# Numerical investigations of a cylindrical Hall thruster

**IEPC-2011-070**

K. Matyash[1],
*Max-Planck-Institut für Plasmaphysik, EURATOM Association, Greifswald, D-17491, Germany*

R. Schneider[2],
*Greifswald University, Greifswald, D-17487, Germany*

O. Kalentev[3]
*Max-Planck-Institut für Plasmaphysik, EURATOM Association, Greifswald, D-17491, Germany*

Y. Raitses[4], N. J. Fisch[5],
*Princeton Plasma Physics Laboratory, Princeton, New Jersey 08543, USA*

**Abstract:** A new 3 dimensional Particle-in-Cell code was applied to simulate the operation of a 100 W CHT thruster. In the simulation a spoke moving with a velocity of about 0.8 cm/µs was observed. The initial position of the spoke was found to be strongly correlated with the cathode placement. The simulation has shown that the depletion of neutral gas can lead to azimuthal asymmetry of the discharge and possibly to the spoke phenomenon.

## Nomenclature

| | | |
|---|---|---|
| $B$ | = | magnetic field |
| $e$ | = | electron charge |
| $E$ | = | electric field |
| $\dot{m}$ | = | mass flow rate |
| $n_e$ | = | electron density |
| $T_e$ | = | electron temperature |
| $Q$ | = | electric charge |
| $\Delta t$ | = | time step |
| $v_e$ | = | electron velocity |
| $U$ | = | voltage |
| $\Delta x$ | = | cell size |
| $X, Y, Z$ | = | coordinates |
| $\varepsilon$ | = | dielectric permittivity |
| $\varepsilon_0$ | = | vacuum permittivity |
| $\varphi$ | = | electrostatic potential |
| $\rho$ | = | charge density |

[1] Research Physicist, Max-Planck-Institute for Plasma Physics, konstantin.matyash@ipp.mpg.de
[2] Professor, Greifswald University, Head of Computer Center, schneider@uni-greifswald.de
[3] Research Physicist, Max-Planck-Institute for Plasma Physics, okalenty@ipp.mpg.de
[4] Research Physicist, PPPL, yraitses@pppl.gov
[5] Professor, Princeton University and PPPL, fisch@princeton.edu

## I. Introduction

The Hall thruster (HT)[1] is an electromagnetic propulsion device that uses a cross-field plasma discharge to accelerate ions. The thrust is a reaction force to this acceleration, exerted upon the thruster magnetic circuit. In a conventional HT, axial electric and radial magnetic fields are applied in an annular channel. The magnetic field is large enough to lock the electrons in the azimuthal **E×B** drift, but small enough to leave the ion trajectories almost unaffected. Because of the reduced electron mobility across the magnetic field, a substantial axial electric field can be maintained in the quasi-neutral plasma and electrons can effectively ionize neutral atoms of the propellant gas. Under such conditions, the electric field supplies energy mainly to accelerate the unmagnetized ions. Unlike the space-charge limited gridded ion engine, the HT accelerates the ions in the quasi-neutral plasma. Thus, larger ion current and thrust densities can be achieved.

The drawback of the annular-geometry of a conventional HT is that it has an unfavorable ratio of the channel surface area to the channel volume. The plasma therefore tends to interact with the thruster channel walls, which results in heating and erosion of parts of the thruster[2]. This tendency becomes more pronounced when the HT is scaled down to low power[3].

A cylindrical geometry Hall thruster (CHT)[4], which was proposed and developed at the PPPL has a reduced ratio of the channel surface area to volume which is limiting electron transport and ion losses[5,6]. Initially developed for a kilowatt power level[4], this non-conventional HT concept was extensively studied and developed for micro propulsion applications[5-7]. The CHTs demonstrated performance comparable to the state-of-the art annular HTs of similar power levels[4,8].

In our previous work[9] we applied a cylindrically symmetrical 2-dimensional Particle-in-Cell code with Monte Carlo Collisions (PIC MCC)[10] to simulate the operation of the PPPL 100 W cylindrical Hall thruster (CHT)[4,7]. Although, the simulation results were in qualitative agreement with probe measurements of potential, electron temperature and plasma density[11], the simulation was not able to reproduce the dependence of the plasma parameters on the enhancement of cathode electron current observed in the experiment[11]. The possible reason for discrepancy is that in the simulation azimuthal dynamics was missing and anomalous electron transport perpendicular to magnetic field was included in the simulation only empirically via Bohm-type diffusion with fixed diffusion coefficient.

In order to resolve the electron cross-field transport self-consistently, accounting for plasma azimuthal dynamics, we have developed a new fully three-dimensional PIC MCC code. The first results of the 3D code for 100 W CHT are presented in this paper.

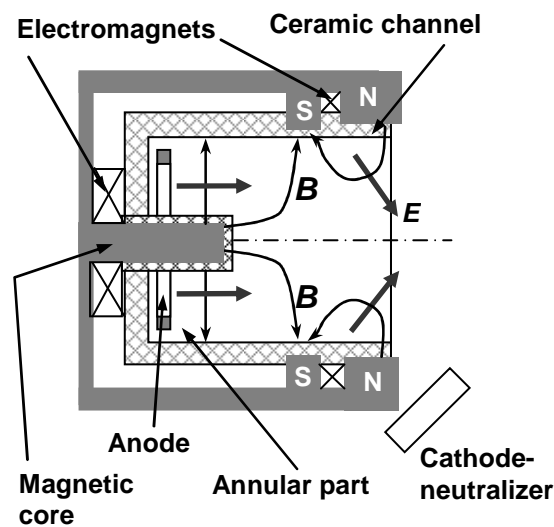

**Figure 1. Schematic of a cylindrical Hall thruster**

The paper is organized as follows: in chapter 2 the main features of cylindrical Hall thruster are outlined. In chapter 2 the PIC MCC model and the simulation results for CHT are presented. Chapter 4 gives a summary and discusses the outlook for future work.

## II. Cylindrical Hall thruster

Fig. 1 illustrates the principle of operation of the cylindrical thruster. A cylindrical Hall thruster consists of a cylindrical ceramic channel, a ring-shaped anode, which serves also as a gas distributor, a magnetic core and magnetized sources (Fig. 1). The magnetic field lines intersect the ceramic channel walls. The electron drifts are closed, with the magnetic field lines forming equipotential surfaces, with $\mathbf{E} = -\mathbf{v}_e \times \mathbf{B}$, where E is the electric field and $v_e$ is the electron drift velocity. The radial component of the magnetic field crossed with the azimuthal electron current produces the thrust. However, the electrons are not confined to an axial position; rather they bounce over an axial region, impeded from entering the annular part of the channel because of magnetic mirroring. Two magnetized sources, electromagnetic coils with opposite currents, can produce a cusp-like magnetic field in the channel, with a strong radial component. To maintain ionizing collisions, the anode (gas inlet) is placed in the short annular part of the channel. The length of the annular part of the channel is designed to minimize the ionization mean free path, thus localizing the ionization of the working gas at the boundary of the annular and cylindrical regions. Hence, most of the voltage drop occurs in the cylindrical region that has large volume-to-surface ratio. This conclusion is supported by the results of plasma measurements[4,7] and by Monte-Carlo simulations[6]. We found that in order to explain the observed discharge current, the electron anomalous collision frequency $v_B$ has to be on the order of the Bohm value, $v_B \approx \omega_c/16$, where $\omega_c$ is the electron gyrofrequency[6].

## III. The model

The detailed description of the PIC-MCC method can be found in thorough reviews[12,13]. Here, we just outline the main features of our model. In PIC-MCC simulations we follow the kinetics of so-called "Super Particles" (each of them representing many real particles), moving in the self consistent electric field calculated on a spatial grid from the Poisson equation. The particle collisions are handled by Monte-Carlo collision (MCC) routines, which randomly change particle velocities according to the actual collision dynamics. The simulation includes electrons, Xe$^+$ ions and the neutral Xenon atoms. All relevant collisional processes are included in the model: electron-neutral elastic, ionization and excitation collisions, ion-neutral momentum-transfer and charge exchange collisions. The dynamics of the background neutral gas is self-consistently resolved with direct simulation Monte Carlo.

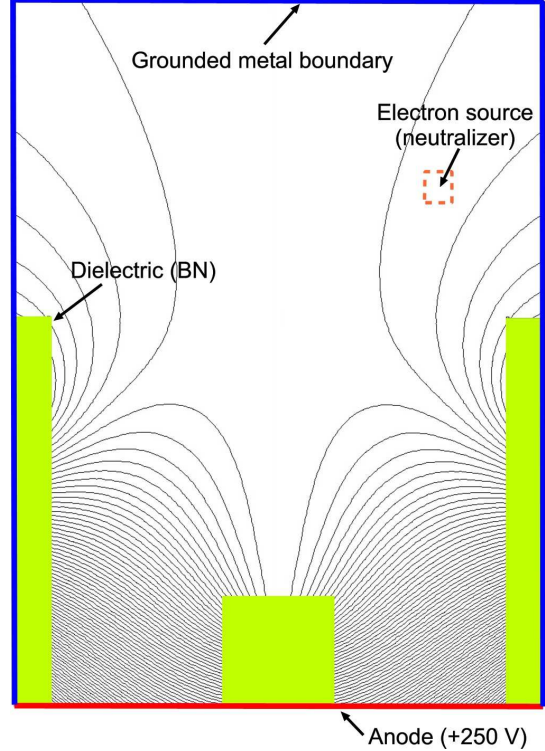

In our model we resolve 3 spatial and 3 velocity components. The model utilizes an equidistant Cartesian grid which explicitly assures momentum conservation and zero self forces. The computational domain in the present simulations is extended beyond the discharge channel and includes the near-field region of the thruster.

The computational domain represents a parallelepiped with length $Z_{max} = 40\ mm$ and sides $X_{max} = Y_{max} = 30\ mm$. Z axis is directed along the thruster symmetry axis. The axial cross-section of the computational domain together with magnetic field topology is shown in Fig. 2. The thruster channel length (distance from the anode to the exit plane) is $Z_{thr} = 22$ mm. The thruster channel walls are located at $X_{ch1} = 3\ mm$, $X_{ch2} = 27\ mm$, $Y_{ch3} = 3\ mm$, $Y_{ch4} = 27\ mm$. At $Z = 0$ the metal anode is located. At the center of the anode the annular part with the length $Z_{anular} = 6\ mm$ and sides $X_{anular} = Y_{anular} = 6\ mm$ is placed. The region outside the thruster exit: ($Z_{thr} < Z < Z_{max}$, $X_{ch1} < X < X_{ch2}$, $Y_{ch1} < Y < Y_{ch2}$) represents the very near-field plume zone.

**Figure 2. Computational domain with magnetic field topology.**

The thruster channel and the annular part are dielectric (Boron Nitride). The lateral and axial boundaries of the computational domain are assumed to be metallic. All metal elements in the simulation, except for the anode are at ground potential. At the anode, a voltage of $U_a = 250$ V is applied.

In order to compute the electrostatic potential in the system, the Poisson equation:

$$\frac{\partial}{\partial x}\varepsilon\frac{\partial \varphi}{\partial x} + \frac{\partial}{\partial y}\varepsilon\frac{\partial \varphi}{\partial y} + \frac{\partial}{\partial z}\varepsilon\frac{\partial \varphi}{\partial z} = -\frac{1}{\varepsilon_0}\rho \qquad (1)$$

is discretized on the grid, taking into account the possible change of the dielectric permittivity $\varepsilon$ across the dielectric surfaces. Applying the centered five-point scheme and assuming that at the point ($x_i, y_j, z_k$) the dielectric surface can be perpendicular to the X, Y or Z direction for the equidistant square grid ($\Delta x_i = \Delta y_j = \Delta z_k = \Delta x$) we get the finite difference equation:

$$\varepsilon_{i-\frac{1}{2}}\varphi_{i-1,j,k} + \varepsilon_{i+\frac{1}{2}}\varphi_{i+1,j,k} + \varepsilon_{j-\frac{1}{2}}\varphi_{i,j-1,k} + \varepsilon_{j+\frac{1}{2}}\varphi_{i,j+1,k} + \varepsilon_{k-\frac{1}{2}}\varphi_{i,j,k-1} + \varepsilon_{k+\frac{1}{2}}\varphi_{i,j,k+1} +$$
$$\left(\varepsilon_{i-\frac{1}{2}} + \varepsilon_{i+\frac{1}{2}} + \varepsilon_{j-\frac{1}{2}} + \varepsilon_{j+\frac{1}{2}} + \varepsilon_{k-\frac{1}{2}} + \varepsilon_{k+\frac{1}{2}}\right)\varphi_{i,j} = -\frac{1}{\varepsilon_0} \cdot \frac{Q^V_{i,j,k} + Q^S_{i,j,k}}{\Delta x} \quad (2)$$

Here

$$\varepsilon_{i-\frac{1}{2}} = \frac{1}{4}\left(\varepsilon_{i-\frac{1}{2},j-\frac{1}{2},k-\frac{1}{2}} + \varepsilon_{i-\frac{1}{2},j-\frac{1}{2},k+\frac{1}{2}} + \varepsilon_{i-\frac{1}{2},j+\frac{1}{2},k-\frac{1}{2}} + \varepsilon_{i-\frac{1}{2},j+\frac{1}{2},k+\frac{1}{2}}\right), \quad (3)$$

$$\varepsilon_{i+\frac{1}{2}} = \frac{1}{4}\left(\varepsilon_{i+\frac{1}{2},j-\frac{1}{2},k-\frac{1}{2}} + \varepsilon_{i+\frac{1}{2},j-\frac{1}{2},k+\frac{1}{2}} + \varepsilon_{i+\frac{1}{2},j+\frac{1}{2},k-\frac{1}{2}} + \varepsilon_{i+\frac{1}{2},j+\frac{1}{2},k+\frac{1}{2}}\right),$$

$$\varepsilon_{j-\frac{1}{2}} = \frac{1}{4}\left(\varepsilon_{i-\frac{1}{2},j-\frac{1}{2},k-\frac{1}{2}} + \varepsilon_{i-\frac{1}{2},j-\frac{1}{2},k+\frac{1}{2}} + \varepsilon_{i+\frac{1}{2},j-\frac{1}{2},k-\frac{1}{2}} + \varepsilon_{i+\frac{1}{2},j-\frac{1}{2},k+\frac{1}{2}}\right),$$

$$\varepsilon_{j+\frac{1}{2}} = \frac{1}{4}\left(\varepsilon_{i-\frac{1}{2},j+\frac{1}{2},k-\frac{1}{2}} + \varepsilon_{i-\frac{1}{2},j+\frac{1}{2},k+\frac{1}{2}} + \varepsilon_{i+\frac{1}{2},j+\frac{1}{2},k-\frac{1}{2}} + \varepsilon_{i+\frac{1}{2},j+\frac{1}{2},k+\frac{1}{2}}\right),$$

$$\varepsilon_{k-\frac{1}{2}} = \frac{1}{4}\left(\varepsilon_{i-\frac{1}{2},j-\frac{1}{2},k-\frac{1}{2}} + \varepsilon_{i-\frac{1}{2},j+\frac{1}{2},k-\frac{1}{2}} + \varepsilon_{i+\frac{1}{2},j-\frac{1}{2},k-\frac{1}{2}} + \varepsilon_{i+\frac{1}{2},j+\frac{1}{2},k-\frac{1}{2}}\right),$$

$$\varepsilon_{k+\frac{1}{2}} = \frac{1}{4}\left(\varepsilon_{i-\frac{1}{2},j-\frac{1}{2},k+\frac{1}{2}} + \varepsilon_{i-\frac{1}{2},j+\frac{1}{2},k+\frac{1}{2}} + \varepsilon_{i+\frac{1}{2},j-\frac{1}{2},k+\frac{1}{2}} + \varepsilon_{i+\frac{1}{2},j+\frac{1}{2},k+\frac{1}{2}}\right),$$

$\varepsilon_{i\pm\frac{1}{2},j\pm\frac{1}{2},k\pm\frac{1}{2}}$ is the dielectric permittivity inside corresponding cell; $Q^V_{i,j,k}$ and $Q^S_{i,j,k}$ are volume and surface charges associated with the grid point. The dielectric permittivity is set to $\varepsilon = 4$ inside the dielectric material (BN) and $\varepsilon = 1$ inside the discharge channel and the near-field region.

If ($x_i, y_j, z_k$) belongs to the metal element, then

$$\varphi_{i,j,k} = U_{i,j,k} \quad . \quad (4)$$

Here $U_{i,j,k} = 250$ V for the anode and $U_{i,j,k} = 0$ for the grounded metal parts. The set of finite difference equations Eq.2, Eq.4 is solved using Watson Sparse Matrix Package[14]. This approach allows us to calculate the potential inside the computational domain, self-consistently resolving the floating potential on the dielectric surfaces.

The particles' equations of motion are solved for discrete time steps with the leap-frog / Boris algorithm[12]. All surfaces in the simulation are assumed to be absorbing for electrons and ions. No secondary electron emission is assumed in the simulation. The neutrals are re-launched from the surfaces using a Maxwellian distribution with temperature $T_n = 1000$ K.

The neutrals are injected into the system through two coaxial ring slits at the anode with the mass flow rate $\dot{m} = 0.4$ mg/s.

Electrons with a Maxwellian distribution and a temperature $T_e = 5$ eV are introduced into the system in the source region $51\,mm < X < 53\,mm$, $13\,mm < Y < 17\,mm$, $26\,mm < Z < 27\,mm$ with uniform density and the constant current $I_c = 0.25$ A, simulating the thruster cathode. In the simulation the electrons from the source, accelerated in the thruster's electric field, are ionizing the neutral gas, creating the plasma in the thruster channel. In order to speed up the plasma ignition process, the supplementary ring source of the electrons with temperature $T_e = 20$ eV and current $I_c = 0.05$ was applied in the annular part during the first 300 ns of the simulation. After that the supplementary source was switched off and only the cathode electron source was operating.

To reduce the computational time the size of the system is scaled down by factor of 10. In order to preserve the ratio of the particles mean free paths and the gyroradii to the system length, the collisions cross-sections and the magnetic field are increased by the same factor 10.

In order to speed up the neutral dynamics, the neutrals were injected into the system with velocity 4 times higher than in the experiment, keeping the mass flow rate constant. To compensate for the neutral density decrease, the cross-sections for collisions with neutrals were multiplied by another factor 4. Thus both the neutral transport and the neutral ionization rate $\frac{\dot{n}_{Xe}}{n_{Xe}}$ in the simulation are accelerated by factor 4.

An equidistant computational grid 60x60x80 was used in the simulation. The total number of computational particles in the simulation was about 30000000. The cell size $\Delta x = \Delta y = \Delta z = 5 \cdot 10^{-2}$ mm in the simulation was chosen to ensure that it is smaller than the smallest Debye length in the system. The time step was set to $\Delta t = 5.6 \cdot 10^{-12}$ s in order to resolve the electron plasma frequency. The simulation was carried on a 4-processor Intel Xeon workstation. The duration of the run was about 10 days. About ~ $10^6$ time steps were performed which corresponds to a simulated time of 6 μs.

The results of the simulations are summarized in Figs. 3-8. At each of the figures snapshots of the electron density, the neutral density and the plasma potential, taken at lateral cross-sections at six axial positions, are presented. The $Z = 2\,mm$ and $Z = 5\,mm$ correspond to the annular part of the thruster channel. Points $Z = 8\,mm$, $Z = 11\,mm$ and $Z = 14.5\,mm$ are in the cylindrical part of the thruster channel, $Z = 14.5\,mm$ corresponds to the position of the cusp at the channel wall. The cross-section at $Z = 26\,mm$ is taken outside of the thruster channel at axial position where the cathode electron source is located.

In Fig. 3 one can see the parameters at $t = 2.691 \cdot 10^{-7}$ s of the simulation. This corresponds to the initial phase, when the supplementary electron source is active in the annular part of the thruster channel. At this phase the electron and the neutral densities in the channel are roughly axially-symmetric. The asymmetry in the potential is caused by the rectangular boundaries. In Fig. 4 at $t = 4.934 \cdot 10^{-7}$ s the supplementary electron source is already switched off, so the electrons are supplied into the system only from the cathode at the thruster exit. By this time a pronounced axial asymmetry in the electron density is formed in the thruster channel. In the annular channel a well localized plasma "spoke" appears. Rotating spokes in the plasma density were observed in numerous Hall thruster experiments[15-21].

The azimuthal position of the spoke is shifted relative to the cathode in direction of the **ExB** drift (clockwise in Fig. 4). This could point to the fact that the spoke position is influenced by the cathode placement. Indeed, in the simulation with same parameters but with a cathode azimuthal position swapped onto the other side (rotated by 180°), the spoke changed its position approximately by 180°. Thus, in our simulation the asymmetry caused by the cathode plays an important role for the origin of the spoke. In the experiment other factors can contribute to the asymmetry as well - neutral gas injection, partial contamination of the anode, asymmetry in magnetic field, etc.

In Fig.5 at $t = 1.691 \cdot 10^{-6}$ s it becomes clearly visible that the spoke causes azimuthal asymmetry in the neutral density profile along the thruster channel, depleting the neutrals due to electron-impact ionization. The depletion of the neutral density at the spoke position reaches ~ 80% at $Z = 11\,mm$. This is also where the electron density reaches its maximum $n_{e\max} = 2.4 \cdot 10^{12}\ cm^{-3}$ along the thruster channel. The rapid ionization in the spoke indicates that a strong electric field is present, heating the electrons. Indeed, the electric field oscillations with frequency $f \approx 10\ MHz$, wavelength $\lambda \approx 4\ mm$ and the amplitude up to 100 V/cm both in the axial and azimuthal directions are observed in the simulation. These fluctuations are responsible for the electron heating and cross-field transport in the simulation. The electron current through the spoke in the simulation was $I_e \approx 0.55\ A$, which is at least by two orders of magnitude larger than current provided by "classic" diffusion due to collisions with neutrals; and no near-wall conductivity is accounted in the model as the secondary electron emission is neglected. The measurements of the electron current flowing through the spoke[20,21] have shown that approximately 50% of the total current is conducted by the spoke. At this point one should mention that oscillations with frequency $f \approx 10\ MHz$ were not observed experimentally in electric field measurements in the CHT[21], although it is not clear whether the frequency limits of the measurement circuit is responsible for this.

In Figs. 6-7 one can see how the spoke changes its azimuthal position towards the region where the neutral density is not depleted. Within ~ 1.2 µs, the spoke moves, rotating about 90° in the direction of **ExB** drift. The average velocity of the spoke is about 0.8 cm/µs, which is a factor ~4 higher than was observed in the experiments[19-21]. Whether the faster spoke motion in the simulation is caused by the accelerated neutral dynamics will be investigated in further simulations. By the time 2.2871 µs the initial depletion of the neutral density (bottom right corner) is completely refilled whereas the new depleted region is created at the current spoke position (bottom left corner). At this time the spoke starts to change its position back to its previous place, but instead of rotating the new spoke is originated in the bottom left corner and the first one is dissolved. Curiously, if one interprets this dynamics as directed motion the corresponding velocity is again about 0.8 cm/µs.

Although the mechanism of the spoke formation and dynamics as well as the anomalous transport inside the spoke is still to be understood, in our simulation we have observed that the spoke motion is connected with the azimuthal depletion of the neutral gas due to ionization inside the spoke. The clarification of the phenomena underlying the spoke formation and the dynamics will be the goal of our further research.

Discussing the simulation results and their relation to the experiment one needs to keep in mind that the simulation time 6 µs in experiment corresponds to a start-up phase of the thruster, whereas the spoke measurements inside the thruster[21] are performed at steady-state operation regime. Longer computations are necessary to be able to make quantitative comparisons with the measurements in the steady state. Recently, fast camera imaging of the 9 cm CHT ignition[22] discovered a strong azimuthal asymmetry in radiation from the channel during the first 10 µs of the thruster operation, which indicates presence of the spoke-like structure inside the channel during the thruster ignition. Probably, the spoke observed in our simulation is relevant to transitional regime observed in the Ref.22.

## IV. Conclusion

A new, fully three-dimensional PIC MCC model was developed. The model was applied to simulate the operation of a 100 W CHT thruster. In the simulation a spoke moving with the velocity of about 0.8 cm/µs was observed, which is a factor ~4 higher than in the experiments. The initial position of the spoke was found to be strongly correlated with the cathode placement. The simulation has shown that the depletion of neutral gas can lead to azimuthal asymmetry of the discharge and possibly to the spoke phenomenon.

The clarification of the phenomena underlying the spoke formation and the dynamics as well as electron transport inside the spoke will be the goal of our further research.

## Acknowledgments

This work was supported by the German Space Agency DLR through Project 50 RS 0804 and the U.S. DOE under Contract DE-AC02-09CH11466.

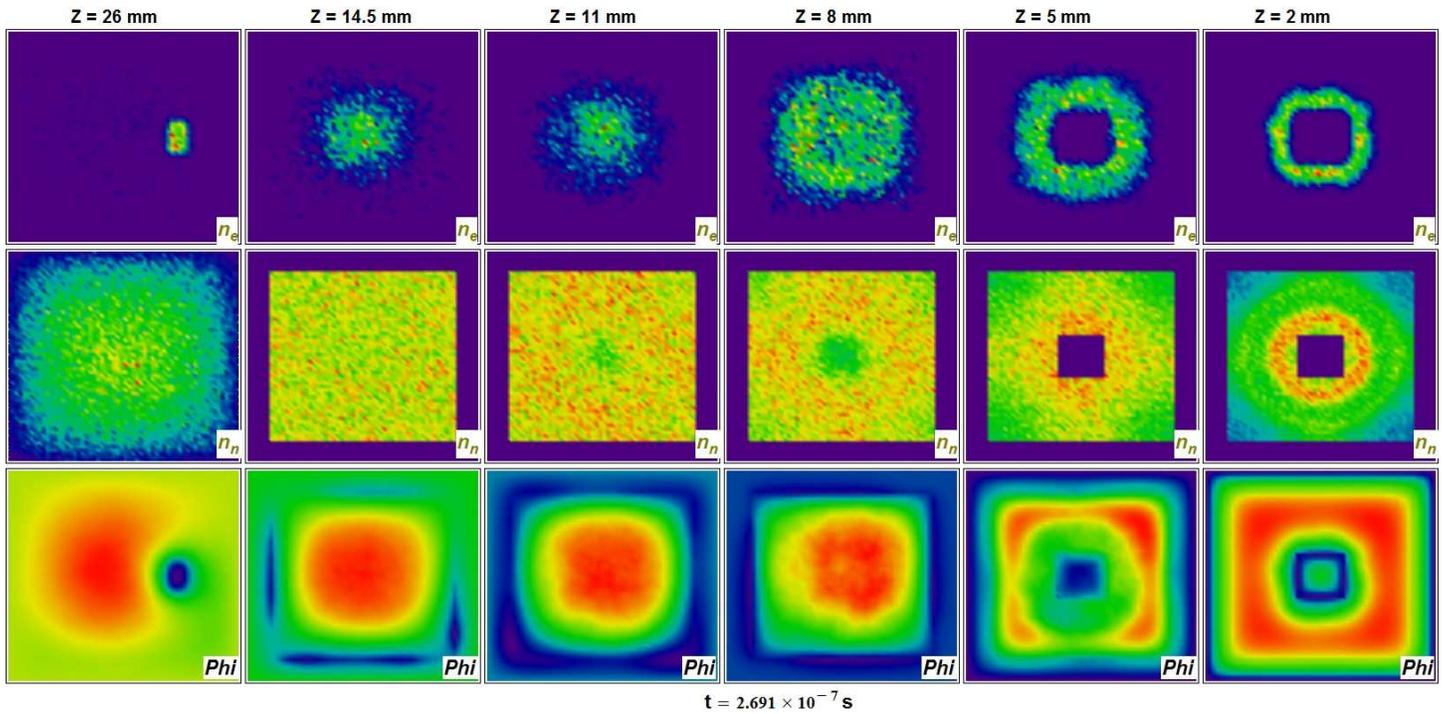

**Figure 3.** Snapshot of the plasma parameters at $t = 2.691 \cdot 10^{-7}$ s taken at six lateral cross-sections. Top row - plasma density, middle row - neutral density, bottom row – plasma potential. All values are given in relative units: reds correspond to the maximum value, blues to the minimum.

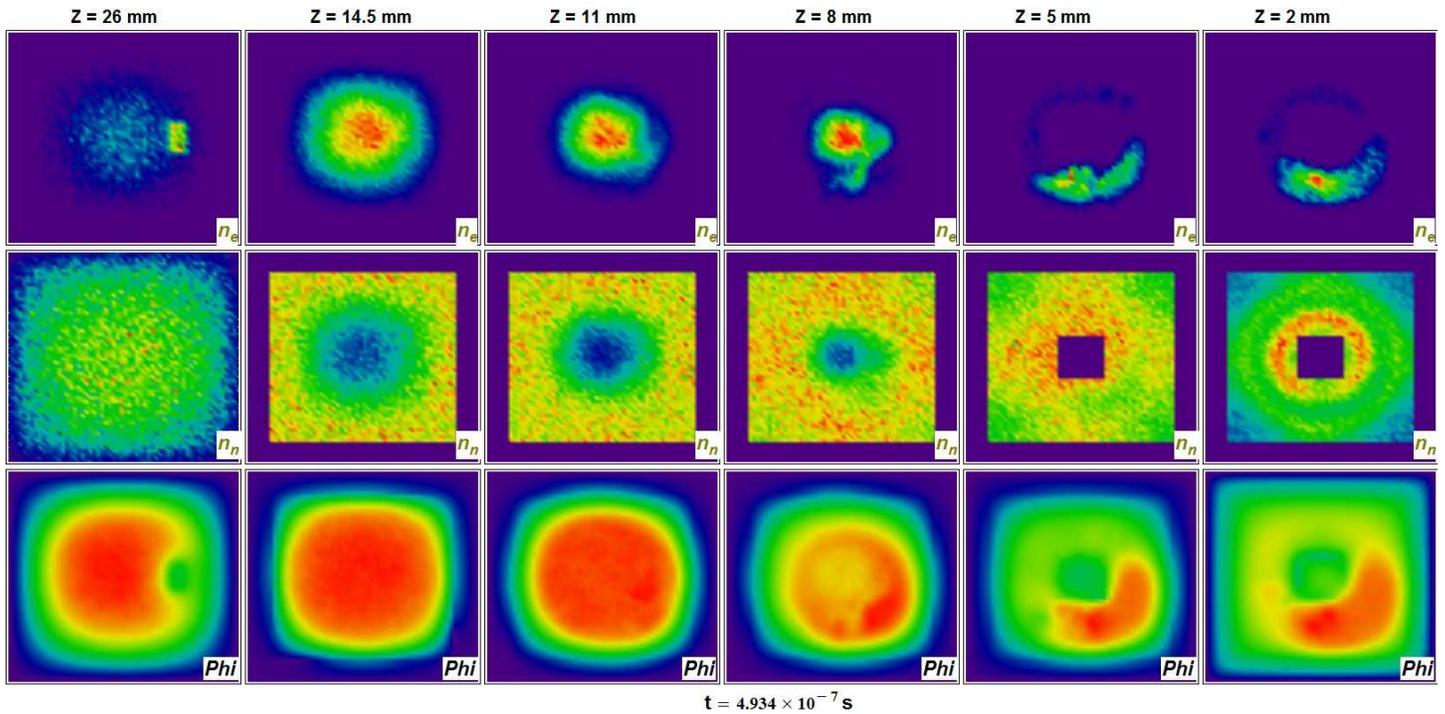

**Figure 4.** Snapshot of the plasma parameters at $t = 4.934 \cdot 10^{-7}$ s taken at six lateral cross-sections. Top row - plasma density, middle row - neutral density, bottom row – plasma potential. All values are given in relative units: reds correspond to the maximum value, blues to the minimum.

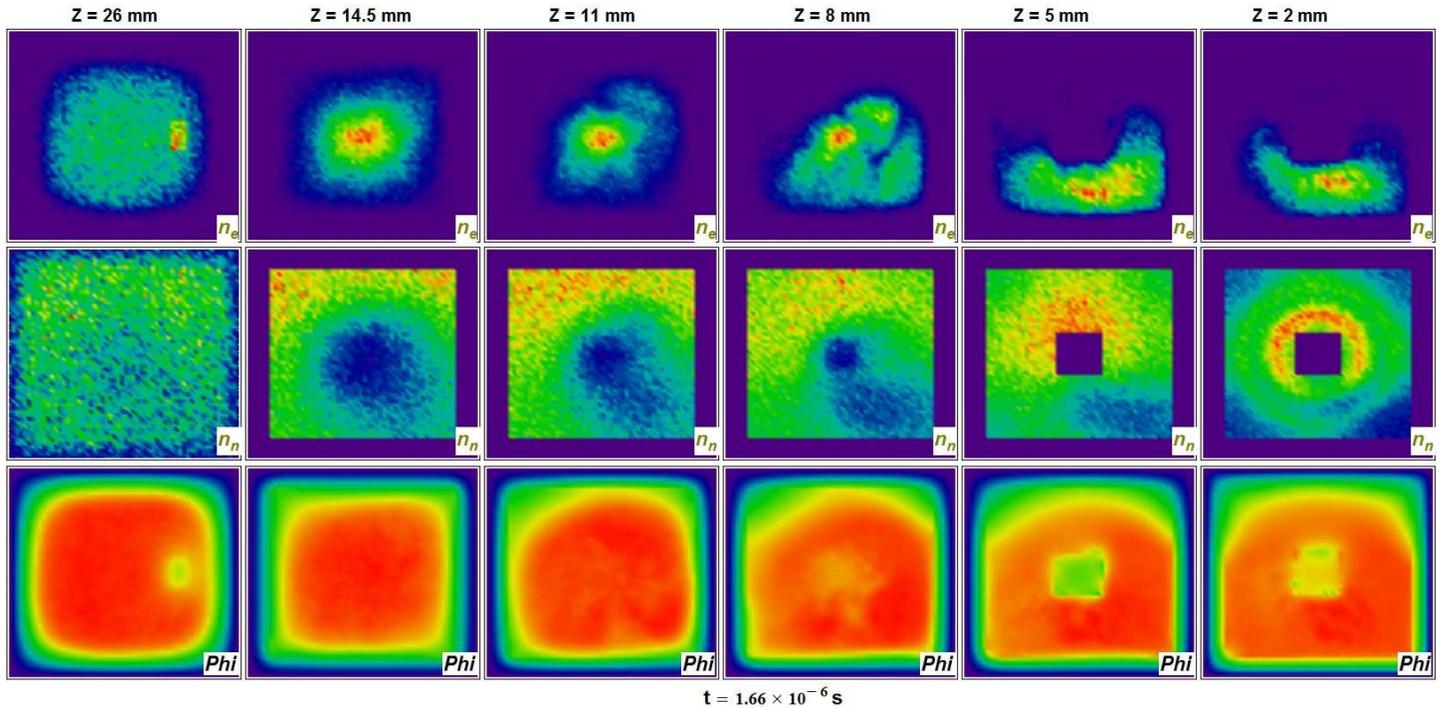

**Figure 5.** Snapshot of the plasma parameters at $t = 1.691 \cdot 10^{-6}$ s taken at six lateral cross-sections. Top row - plasma density, middle row - neutral density, bottom row – plasma potential. All values are given in relative units: reds correspond to the maximum value, blues to the minimum.

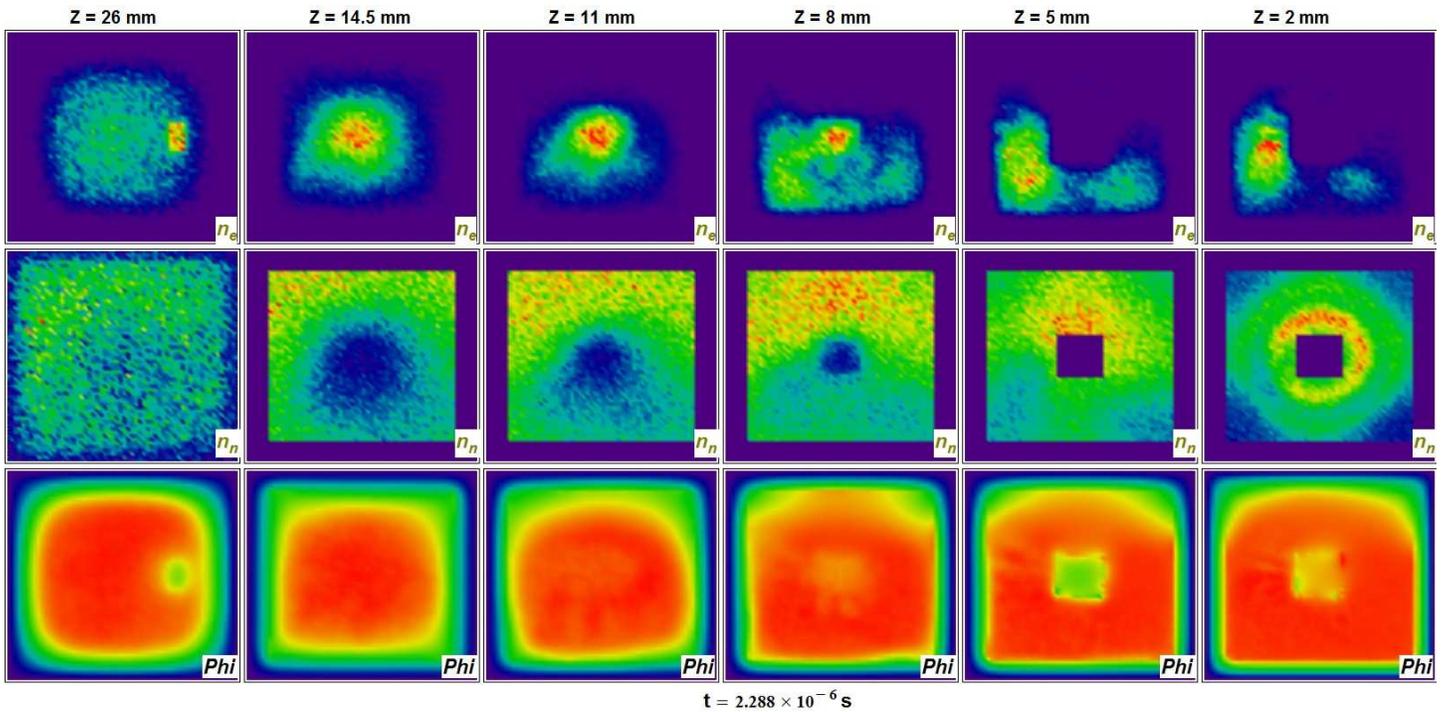

**Figure 6.** Snapshot of the plasma parameters at $t = 2.288 \cdot 10^{-6}$ s taken at six lateral cross-sections. Top row - plasma density, middle row - neutral density, bottom row – plasma potential. All values are given in relative units: reds correspond to the maximum value, blues to the minimum.

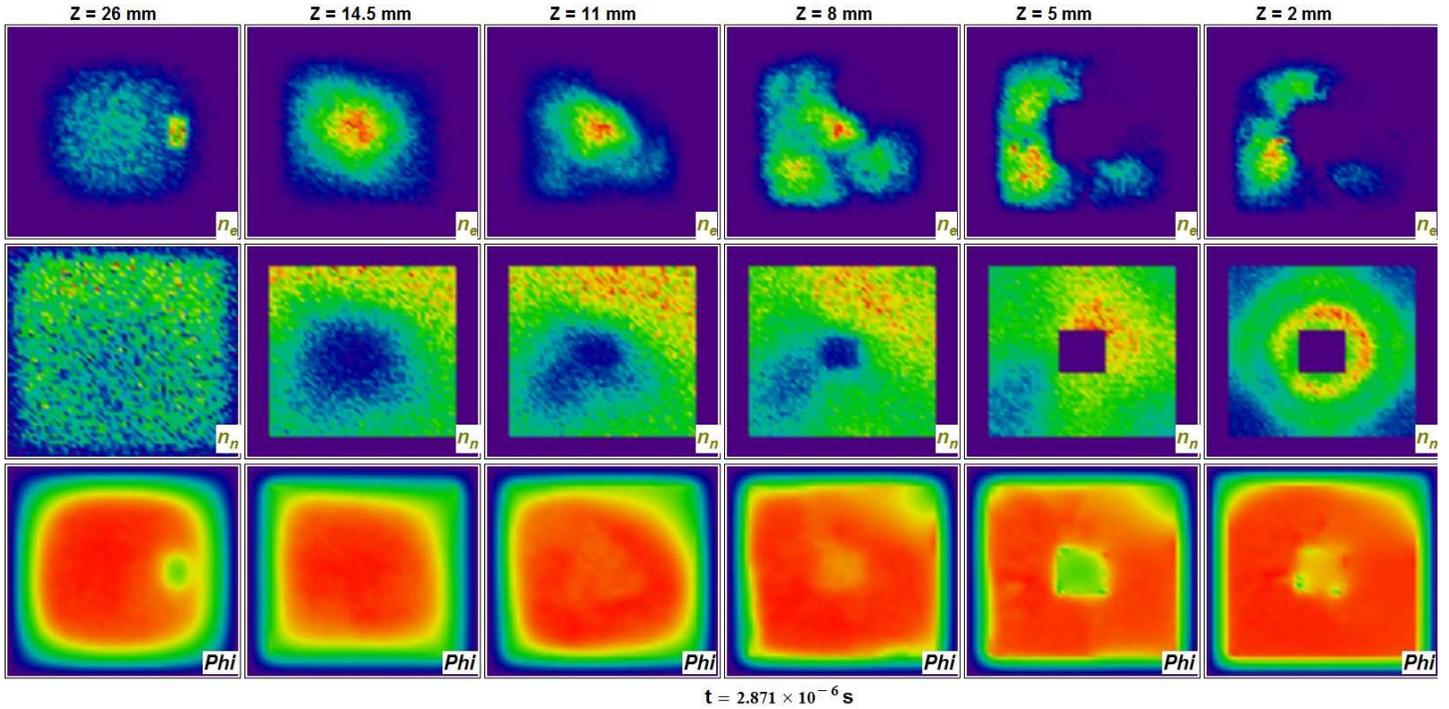

**Figure 7.** Snapshot of the plasma parameters at $t = 2.871 \cdot 10^{-6}$ s taken at six lateral cross-sections. Top row - plasma density, middle row - neutral density, bottom row – plasma potential. All values are given in relative units: reds correspond to the maximum value, blues to the minimum.

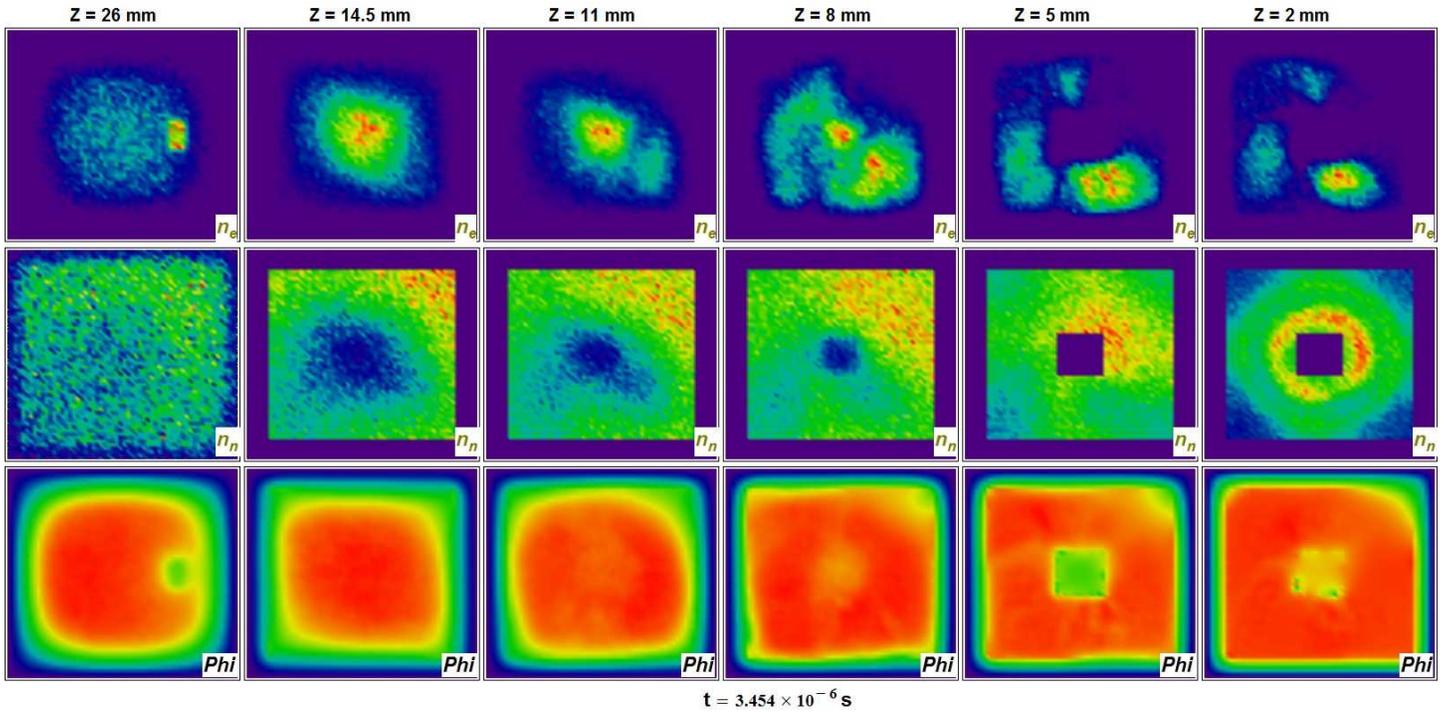

**Figure 8.** Snapshot of the plasma parameters at $t = 4.306 \cdot 10^{-6}$ s taken at six lateral cross-sections. Top row - plasma density, middle row - neutral density, bottom row – plasma potential. All values are given in relative units: reds correspond to the maximum value, blues to the minimum.

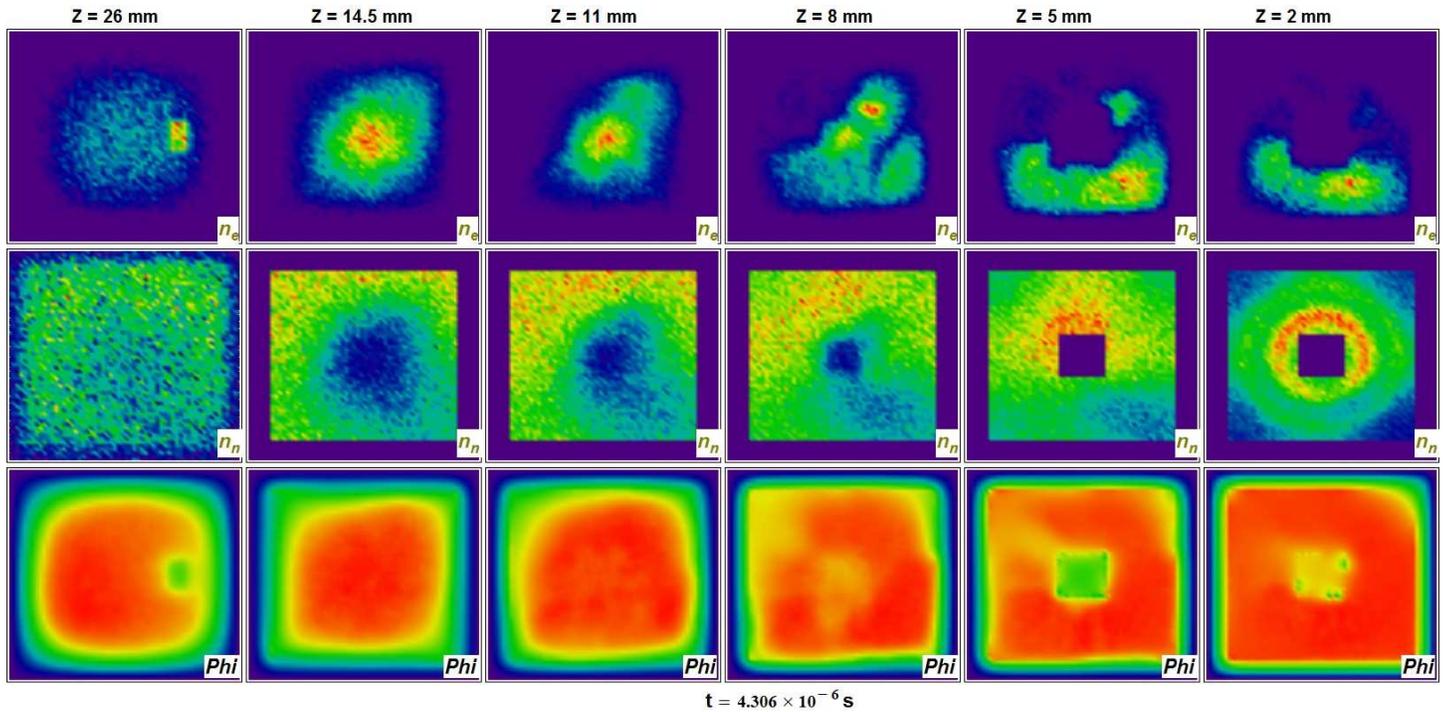

**Figure 9.** Snapshot of the plasma parameters at $t = 4.306 \cdot 10^{-6}$ s taken at six lateral cross-sections. Top row - plasma density, middle row - neutral density, bottom row – plasma potential. All values are given in relative units: reds correspond to the maximum value, blues to the minimum.